\documentstyle[twocolumn,epsf]{article}
\newcommand{\beq}{\begin{equation}}
\newcommand{\eeq}{\end{equation}}
\newcommand{\bea}{\begin{eqnarray}}
\newcommand{\eea}{\end{eqnarray}}

\def\nuc#1#2{\relax\ifmmode{}^{#1}{\protect\text{#2}}\else${}^{#1}$#2\fi}
\newcommand{\dsp}{\displaystyle}
\title{\large
EXPERIMENTAL AND COMPUTER SIMULATION STUDY OF RADIOACTIVITY
OF MATERIALS IRRADIATED BY INTERMEDIATE ENERGY PROTONS}
\vspace{1cm}
\author{Yury E. Titarenko, Oleg V. Shvedov, Vyacheslav F. Batyaev, Evgeny I. 
Karpikhin, \\ Valery M. Zhivun, Ruslan D. Mulambetov\\
{\it Institute for Theoretical and Experimental Physics,
B.Cheremushkinskaya 25,}\\ {\it 117259 Moscow, Russia,
e-mail: Yury.Titarenko@itep.ru} \vspace{0.1cm} \\
Stepan G. Mashnik, Richard E. Prael, William B. Wilson  \\
{\it Los Alamos National Laboratory, Los Alamos, NM 87545, USA}} 
\begin{document}
\maketitle
\noindent ABSTRACT
\vspace{0.1cm}

The results of measurements and computer simulations of radioactivities
and dose rates
as functions of decay time are presented for \nuc{nat}{Pb} and 
\nuc{209}{Bi} irradiated 
by 1.5-GeV protons, \nuc{59}{Co}, \nuc{63}{Cu}, and
\nuc{65}{Cu} irradiated 
by 0.13- and 1.2-GeV 
protons, and \nuc{232}{Th} and \nuc{nat}{U} irradiated by 0.1- and 
0.8-GeV protons. The 
activities and dose rates are measured by direct high-precision 
$\gamma$-spectrometry. The irradiations were made using external beams 
extracted from the ITEP U-10 proton synchrotron. 
Simulations made using the LCS and CINDER'90 code systems are compared 
with measurements.

\vspace{0.4cm}
\noindent INTRODUCTION
\vspace{0.1cm}

The design of a hybrid system driven via a high current accelerator needs 
to have information about residual nuclides that are produced by high 
energy protons in the target and in constructional elements, defining a 
number of technical features of the system. 

Our previous works [1,2] present the results of experimental 
measurements of residual radioactive nuclides produced in the target and 
constructional materials irradiated with 0.13- and 1.5-GeV protons. 
Predictive powers of a number of high energy codes were investigated 
using the experimental data. A general conclusion found insufficient 
accuracy of the codes to simulated radionuclide yields, both independent 
and cumulative, with the mean deviation of simulated results from 
experimental data often of a factor of 2 or higher.  
Reliable information on the yield of each residual nuclide is very 
important and such characteristic can be defined as ``individual".

Unlike our previous works, this study is aimed to investigate  
the predictive power of codes not for individual but for ``integral" 
parameters -- time dependent activation and dose-equivalent rate
(called dose rate hereafter) which are directly dependent on summed 
yield of individual radionuclides. 
For such ``integral" characteristics, one may expect a
much better predictive power of codes than in simulations of 
``individual" parameters. We believe that
the dose rate due to eminations of the $\gamma$-s from
any hybrid system part being irradiated is one of the most important 
working parameters in applications.

Here, the LAHET [3] and
HMCNP Monte Carlo transport
codes were used
together with the code  \linebreak 
CINDER'90 [4] 
to simulate the build-up and decay of the products.

\vspace{0.4cm}
\noindent EXPERIMENT
\vspace{0.1cm}

The irradiations of experimental samples were carried out using 
external beams of the ITEP proton synchrotron. A sandwich of 
experimental and aluminum samples of the same diameters of 10.5 mm was 
placed perpendicularly to
the proton beam during each irradiation. The 
parameters of irradiations and of experimental samples are presented  in 
Table 1.

The CANBERRA spectrometer based on the GC-2518 Ge detector was used 
to measure $\gamma$-spectra of irradiated experimental and aluminum 
samples. $\gamma$-spectra processing and nuclide identification were 
made using the ASPRO and SIGMA codes and the PCNUDAT nuclear database. 
\nuc{24}{Na} formation in aluminum samples was used to determine 
the irradiation 
time integrated proton flux. The details of irradiation techniques, 
schemes and parameters of external beams, techniques of nuclides 
identification and their cross section determination can be found in 
[1].


\begin{table}[ht]\caption{Parameters of irradiations and experimental 
samples}
\vspace{0.2cm}
\begin{tabular}{|p{0.3cm}|p{0.9cm}|p{1.0cm}|p{1.0cm}|p{0.9cm}|p{1.6cm}|}\hline
N&Target&\multicolumn{4}{|c|}{Parameter}\\ \cline{3-6}
&&Sample weight (mg)&Proton energy (MeV)&
Irradi-ation time (min)&Proton flux (p/cm2) during irradiation\\ \hline
A&\nuc{nat}{Pb}&143.8&1478&15&2.43E+13\\
B&\nuc{209}{Bi}&368.05&1478&15&2.67E+13\\
C&\nuc{59}{Co}&202.7&1186&45&6.86E+13\\
D&\nuc{59}{Co}&196.6&127&60&2.29E+13\\
E&\nuc{63}{Cu}&80.2&1186&70&1.08E+14\\
F&\nuc{65}{Cu}&92.6&1186&70&1.37E+14\\
G&\nuc{63}{Cu}&425.55&127&9&1.56E+11\\
H&\nuc{65}{Cu}&521.6&127&9&1.49E+11\\
I&\nuc{232}{Th}&87.5&97&40&6.31E+12\\
J&\nuc{232}{Th}&89.6&795&30&4.86E+12\\
K&\nuc{nat}{U}&159.2&97&60&9.22E+12\\
L&\nuc{nat}{U}&160.9&795&60&7.43E+13\\
\hline\end{tabular}
\vspace{0.4cm}
\end{table}

\vspace{0.4cm}
\noindent SIMULATION TECHNIQUE
\vspace{0.1cm}

The simulation of build-up during irradiations and following decay of 
produced nuclides was made using the LCS [3] 
and 
CINDER'90 [4] code systems.

LCS includes the LAHET and HMCNP Monte-Carlo 
radiation transport codes. LAHET
follows protons of all energies 
and secondary neutrons above 20 MeV throughout the
geometry using a selection of on-line 
nuclear models and parameters; for these 
calculations \linebreak LAHET2.83 was used with default 
selections except for use of
ISABEL intranuclear-cascade model with 
  default parameters,
preequilibrium model following intranuclear
    cascade, and
nuclear elastic scattering for neutrons
    and protons.
 
LAHET passes the position, direction, and
energy of all neutrons below 20 MeV as source
particles to \linebreak HMCNP, which uses evaluated cross-section 
data to follow all neutrons (and 
photons, if desired) throughout the geometry.

The regional nuclide production  
(At./proton) from LAHET and the binned neutron 
fluence (n/bin-p-cm$^2$) from HMCNP are scaled
by region volume (cm$^3$) and proton source
strength (p/s) to form production
probabilities (At./cm$^3$-s) and neutron fluxes
(n/cm$^2$-s). These are used with evaluated
cross sections and decay data in CINDER'90
to calculate the temporal nuclide inventory
during and following irradiation. 

In these simulations of thin-target measurements in environment 
having no features contributing to lower-energy neutron flux,
contributions to nuclide inventory from the lower
energy neutron reactions in HMCNP and in CINDER'90,
although included, are negligible.

To investigate the real predictive power of 
the codes but not just an ability to describe some known 
measured data, 
all simulations have been performed at Los Alamos
strictly before receiving any experimental data 
from Moscow that could be 
compared with theoretical results. 

\vspace{0.4cm}
\noindent PROCESSING EXPERIMENTAL AND SIMULATED DATA
\vspace{0.1cm}

\noindent{
{\em Determination of experimental activities and dose rates}
}

\vspace{0.1cm}

Activity of each nuclide in experimental samples was calculated via 
least square
approximating nuclide counting rates of corresponding 
$\gamma$-energy peaks taking into account $\gamma$-yields and absolute 
effectiveness of the spectrometer at corresponding $\gamma$-energy.
The total activities of the targets were defined as a sum of 
activities of individual nuclides. 
The individual activities were determined dependently of properties 
of precursors present for practically all measured products:

1) 
If precursors are absent or have lifetimes that are either much less 
than the first measurement time after irradiation or much higher 
than the last 
measurement time, the following formula was used for activity calculation:

$$
a(t,T)={A_0 \over \varepsilon \eta} \exp [-\lambda (t-T)] \mbox{ .} 
$$

2) 
If precursors have lifetimes that are comparable with times passed 
after irradiation, the following formula was used:
$$
a(t,T)={1 \over \varepsilon \eta} \{ A_1 \exp [-\lambda_1 (t-T)]
 + A_2 \exp [-\lambda_2 (t-T)]\} \mbox{ ,} 
$$
where,
$A_0$, $A_1$, and $A_2$ are the
coefficients determined via the 
least square
approximation;
$\lambda$, $\lambda_1$, and $\lambda_2$ are the
decay constants of the nuclide and its precursors;
$t$ is the time after starting the irradiation;
$T$ is the irradiation duration;
$\eta$ is the  $\gamma$-yield;
$\varepsilon$ is the absolute effectiveness of the spectrometer.

The dose rate at 1 cm distance from a target was defined via summing 
individual nuclides dose rates which were determined via individual 
activities multiplied by $K_{\gamma}$-dose coefficients 
calculated for each nuclide using 
methods and data described in [5].

\vspace{0.4cm}

\noindent{
{\em Simulated activities and dose rates}
}

\vspace{0.1cm}

To have a correct comparison of obtained experimental results with the
simulated ones, the simulated total activities were calculated by 
summing the individual simulated activities of only the nuclides that were 
identified in measurements and used for calculating experimental 
activities. The simulated dose rates were calculated by summing the 
individual dose rates of the same nuclides.

\vspace{0.5cm}
\noindent AGREEMENT BETWEEN SIMULATED AND EXPERIMENTAL RESULTS 
\vspace{0.1cm}

The determined experimental and simulated activities are presented in 
Figures 1--6. Additionally, the figures show the mean squared deviation 
factors calculated via
$$ 
\langle F \rangle = 10^
{\sqrt{\dsp 
{\Large \langle}
\left(\lg\left( \frac{
\mbox{theor. value}
}{
\mbox{exp. value}
}
\right)\right)^2
{\Large \rangle}}} 
$$
for each of the data sets compared,
where $\langle ...\rangle$ designates averaging over 
all of the comparison events.

As seen from the figures, the simulated results are in good agreement 
with experimental data for most of the targets. In most cases, the 
mean squared deviation factor is from 1.15 to 1.47. The following 
variations of simulated-to-experimental agreements along time range 
are observed:
\begin{enumerate}
\item \nuc{nat}{Pb} and \nuc{209}{Bi}: 
the simulated activities during the first 15 
hours after irradiation are on the average underestimated
by a factor of 2, 
meanwhile, the results are in a much better agreement later.
\item \nuc{63}{Cu}, $E_p = 1186$ MeV: 
the simulated results, both for activity and dose rate, 
are in average 3 times underestimated during the first 5-300 hours after 
irradiation, meanwhile the discrepancies are negligible for other 
times.
\item \nuc{63}{Cu} and \nuc{65}{Cu}, $E_p = 127$ MeV: 
the simulated results, both of activity and dose 
rate, are in average 1.5 times lower 
for \nuc{63}{Cu}, and  2 times underestimated for \nuc{65}{Cu} 
after $\sim$ 1 day after irradiation, meanwhile during 
the first day after irradiation the results either practically 
coincide with experimental data (\nuc{63}{Cu}) or lie very close to them 
(\nuc{65}{Cu}).
\end{enumerate}

The largest discrepancy between simulated and experimental data is 
observed for \nuc{59}{Co} irradiated with 127 MeV protons. 
Simulated activities are about 4 
times, and dose rates -- 2 times underestimated almost for the entire
range of time. This reflects, probably, the importance of nuclear
structure effects for this reaction (relatively low bombarding
energy and vicinity of the target \nuc{59}{Co} to the doubly magic
\nuc{56}{Ni} and magic \nuc{57}{Ni} nuclides) and an inadequate 
capability of LAHET to describe nuclide production near closed shells,
which could be a result of using not good enough approximations
for level densities, for
inverse cross sections of evaporated and preequilibrium particles, 
and/or, for nuclear masses and binding energies near closed shells.

\vspace{0.4cm}

\noindent CONCLUSIONS
\vspace{0.1cm}

As a whole, 
the simulated activities and dose rates are in           
good agreement 
with experimental data for most of the targets along the whole time
range. 
Nevertheless, some serious discrepancies
at certain times after irradiation, especially for \nuc{59}{Co},
were obtained.

At the moment, we have not made a detailed investigation of the 
causes of the observed discrepancies. The list of products that are 
produced in the irradiations and/or are observed at decay phase and 
much contributes to the discrepancies observed have not been defined yet. 
Also, the contribution of various modes of nuclides formation in 
nuclear interactions (spallation, fission, fragmentation) to the observed 
discrepancies has not been analyzed properly.

This work is our first step in the study of experimental and 
simulated activities of targets induced by intermediate and 
high energy protons. 
The analysis of observed discrepancies will be continued, and, 
if possible, will be extended to thick targets in which 
secondary neutrons, rather than the incident protons, have 
the largest contribution to target activation. 
The use of the 
ISABEL model
in LAHET for the higher-energy simulations above 
its nominal 1-GeV limit could lead to inconsistencies;
similar calculations using the alternate
Bertini model,
or preferably 
the nuclear models of the 
CEM95 code [6], 
in LAHET
are called for.
\\

The experimental part of this work have been supported by the
ISTC Project \#839 and the theoretical
study was supported by the U.~S.~Department of Energy
under contract W-7405-Eng-36.

\vspace{0.5cm}
\noindent REFERENCES
\vspace{0.1cm}
\begin{enumerate}

\item 
Yu. E. Titarenko et al., 
{\em Nucl. Instr. and Meth.} {A 414}, 73 (1998).

\vspace*{-0.2cm}
\item 
Yu. E. Titarenko et al., 
{\em Proc. Second Int. Topical Meeting on Nuclear Applications of Accelerator 
Technology (AccApp'98)}, Gatlinburg, TN, USA, September 20-23, 1998,
pp. 164-171 and references therein.

\vspace*{-0.2cm}
\item
R. E. Prael and H. Lichtenstein,
``User guide to LCS: The LAHET Code System,"
Los Alamos National Laboratory Report LA-UR-89-3014 (1989).

\vspace*{-0.2cm}
\item
W. B. Wilson, T. R. England, and K. A. Van Riper, 
``Status of CINDER'90 Codes and Data," 
{\em Proc. Fourth Workshop on Simulating Accelerator Radiation Environments
(SARE4)}, Knoxville, TN, USA, September 14-16, 1998, 
p. 69.

\vspace*{-0.2cm}
\item
N. G. Gusev, V. P. Mashkovich, G. V. Obvintsev, 
``Gamma Irradiation of Radioactive Isotopes and 
Fission Products," Moscow (1958) [in Russian].

\vspace*{-0.2cm}
\item 
S.~G.~Mashnik,~``User Manual~for the Code CEM95",
JINR, Dubna, 1995; \\
http://www.nea.fr/abs/html/iaea1247.html.

\end{enumerate}


\clearpage

\begin{figure*}
\centerline{\epsfxsize 17.5cm \epsffile{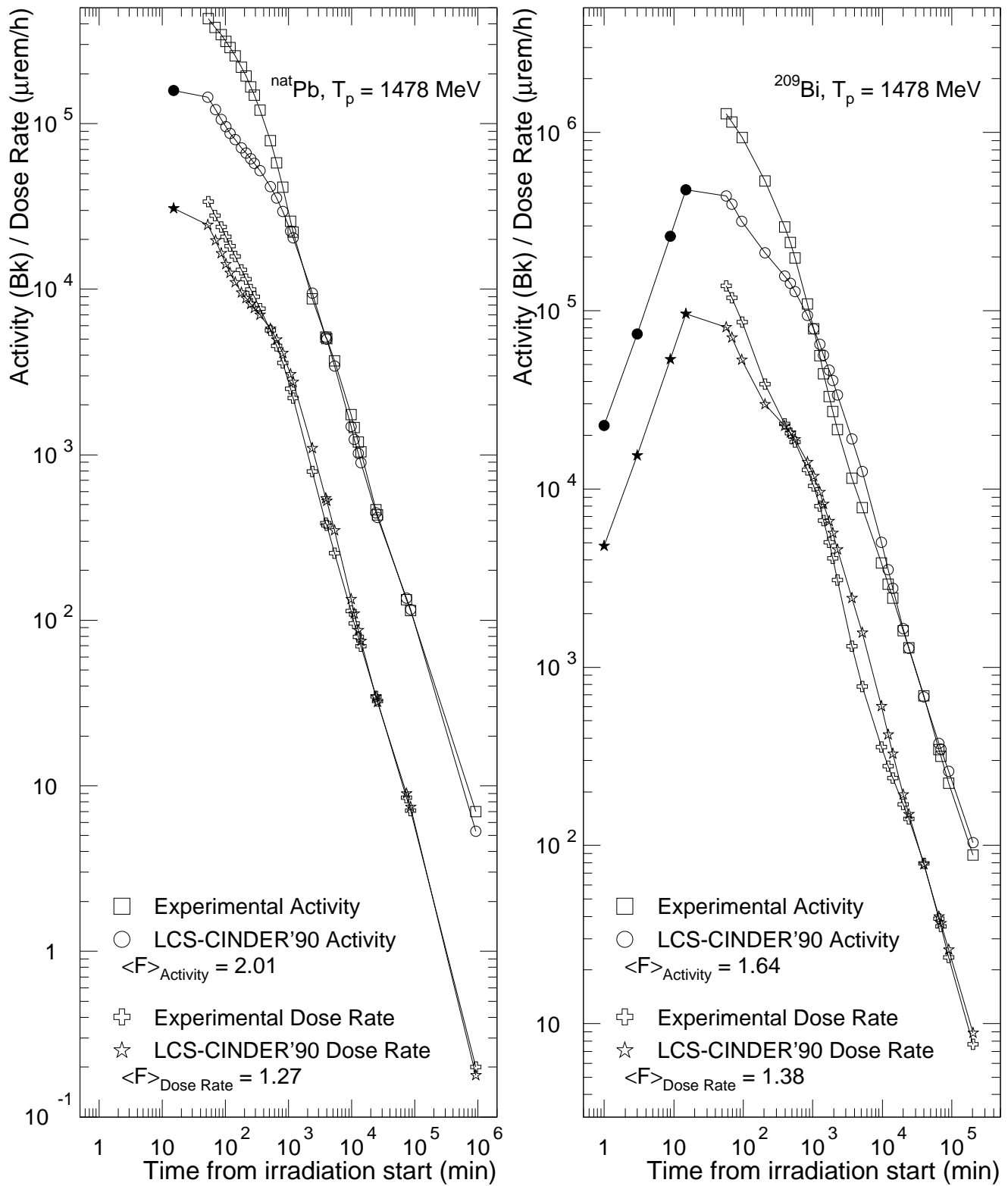}}
\caption{Experimental 
activities and dose rates 
and  results calculated with the LCS-CINDER code
system 
for \nuc{nat}Pb and \nuc{209}{Bi} irradiated with 1478 MeV protons.
The filled symbols show calculated values
during the irradiation.
The mean squared deviation factors (see text) are shown for both.}
\end{figure*}

\begin{figure*}
\centerline{\epsfxsize 17.5cm \epsffile{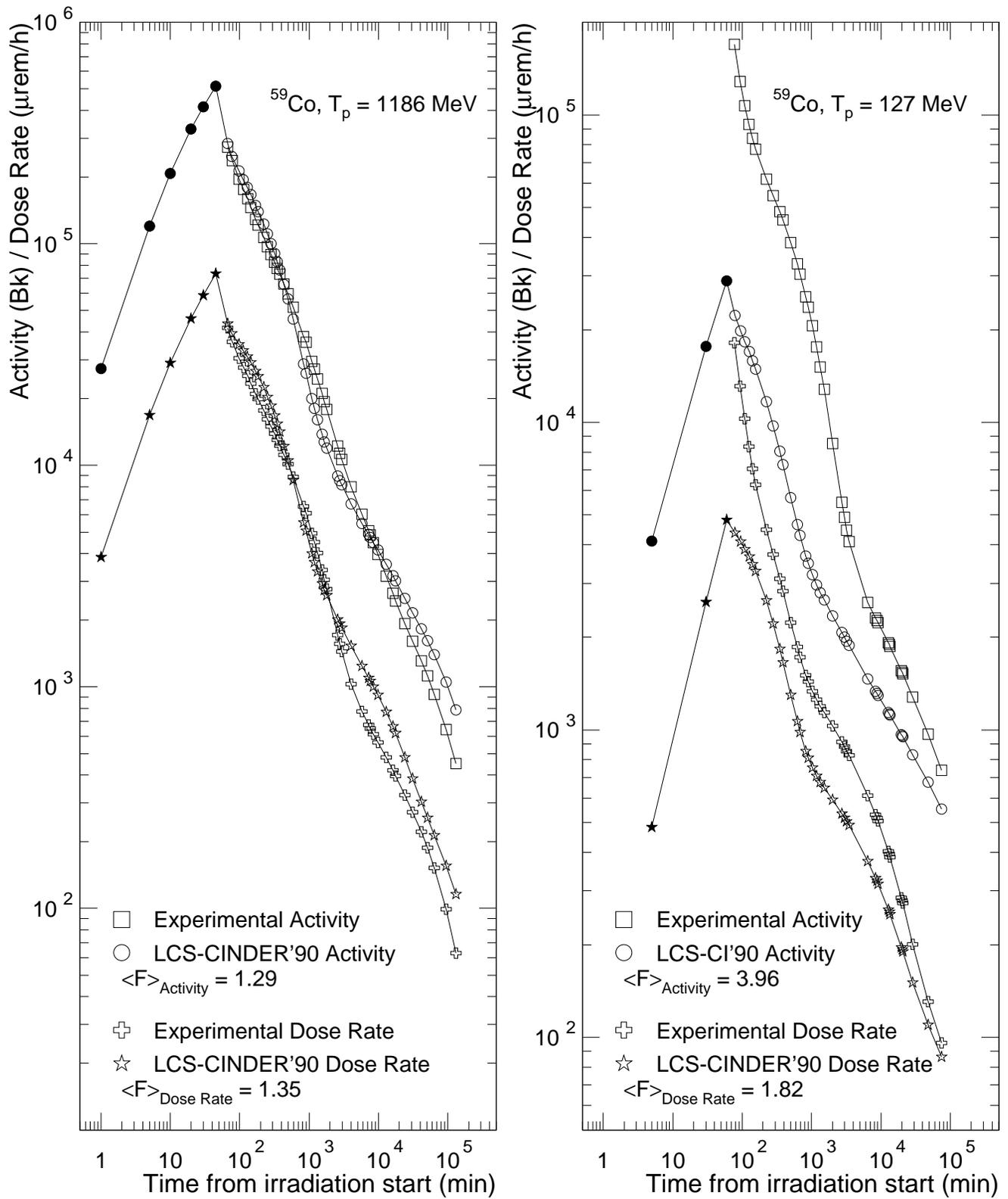}}
\caption{The same as Fig. 1, but for \nuc{59}{Co} irradiated with 1186 MeV 
and 127 MeV protons.}
\end{figure*}

\begin{figure*}
\centerline{\epsfxsize 17.5cm \epsffile{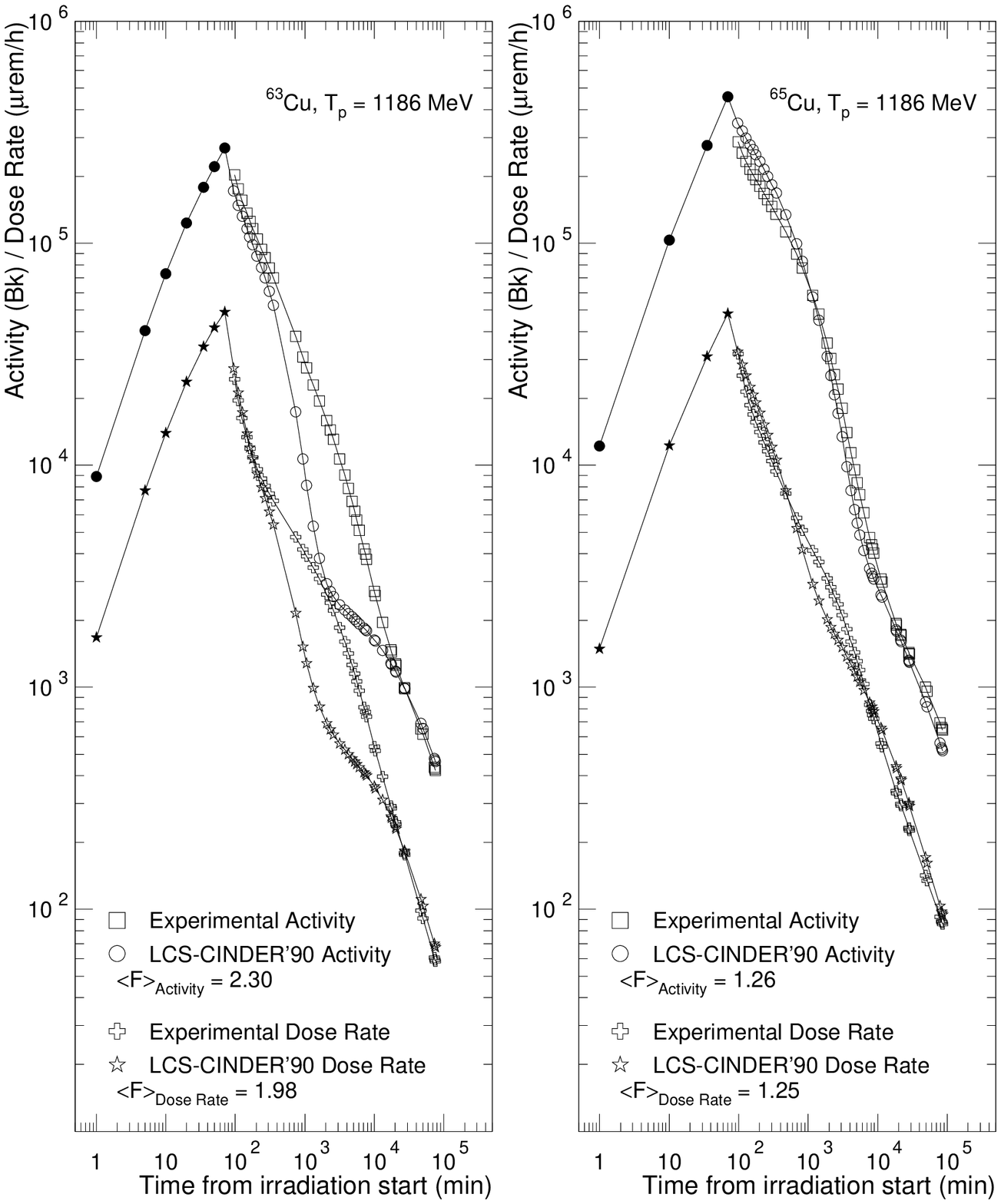}}
\caption{The same as Fig. 1, but for \nuc{63}{Co} and \nuc{65}{Co}
irradiated with 1186 MeV protons.}
\end{figure*}

\begin{figure*}
\centerline{\epsfxsize 17.5cm \epsffile{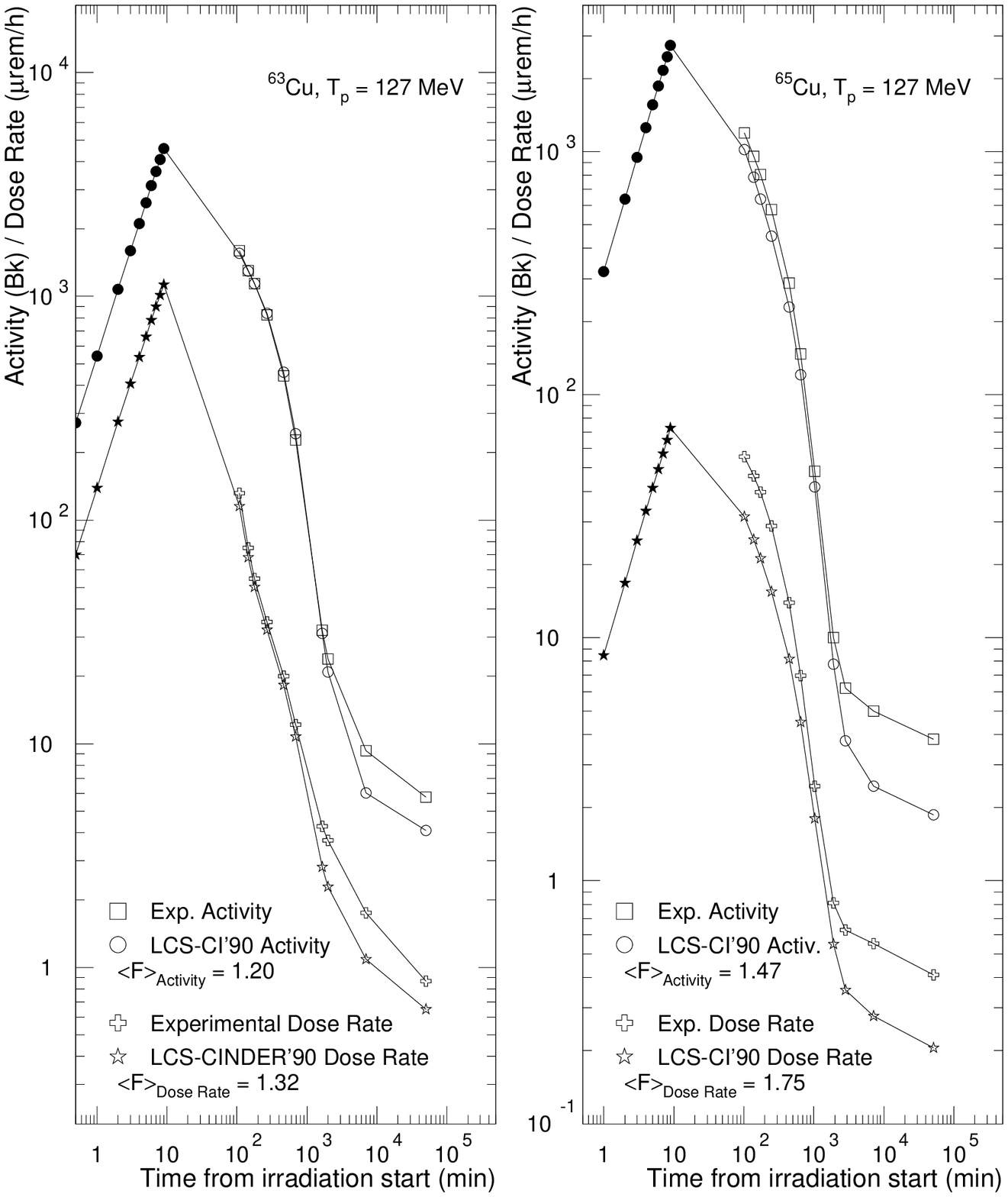}}
\caption{The same as Fig. 1, but for \nuc{63}{Co} and \nuc{65}{Co}
irradiated with 127 MeV protons.}
\end{figure*}

\begin{figure*}
\centerline{\epsfxsize 17.5cm \epsffile{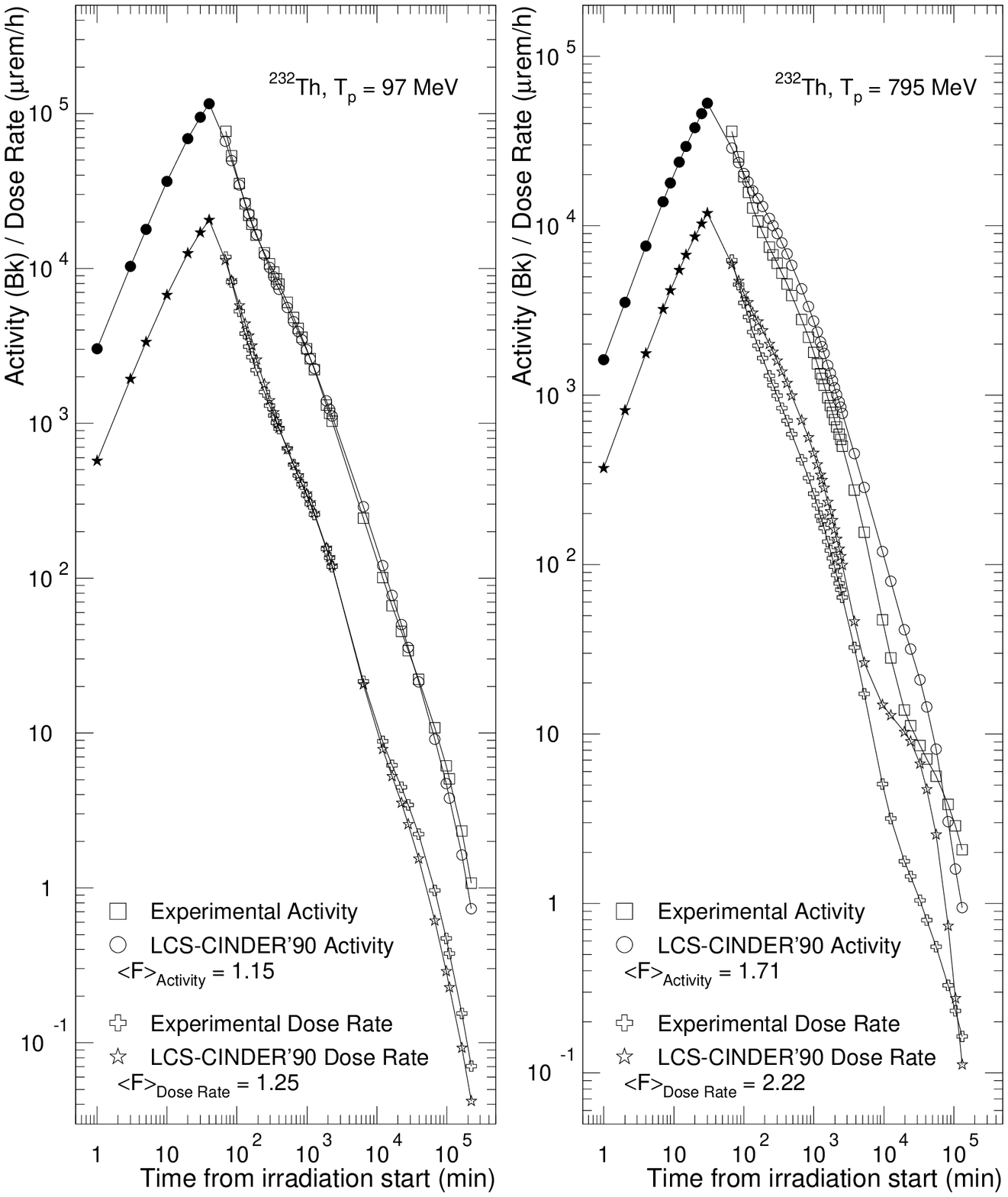}}
\caption{The same as Fig. 1, but for \nuc{232}{Th} irradiated with 97 MeV and 795
MeV protons.}
\end{figure*}

\begin{figure*}
\centerline{\epsfxsize 17.5cm \epsffile{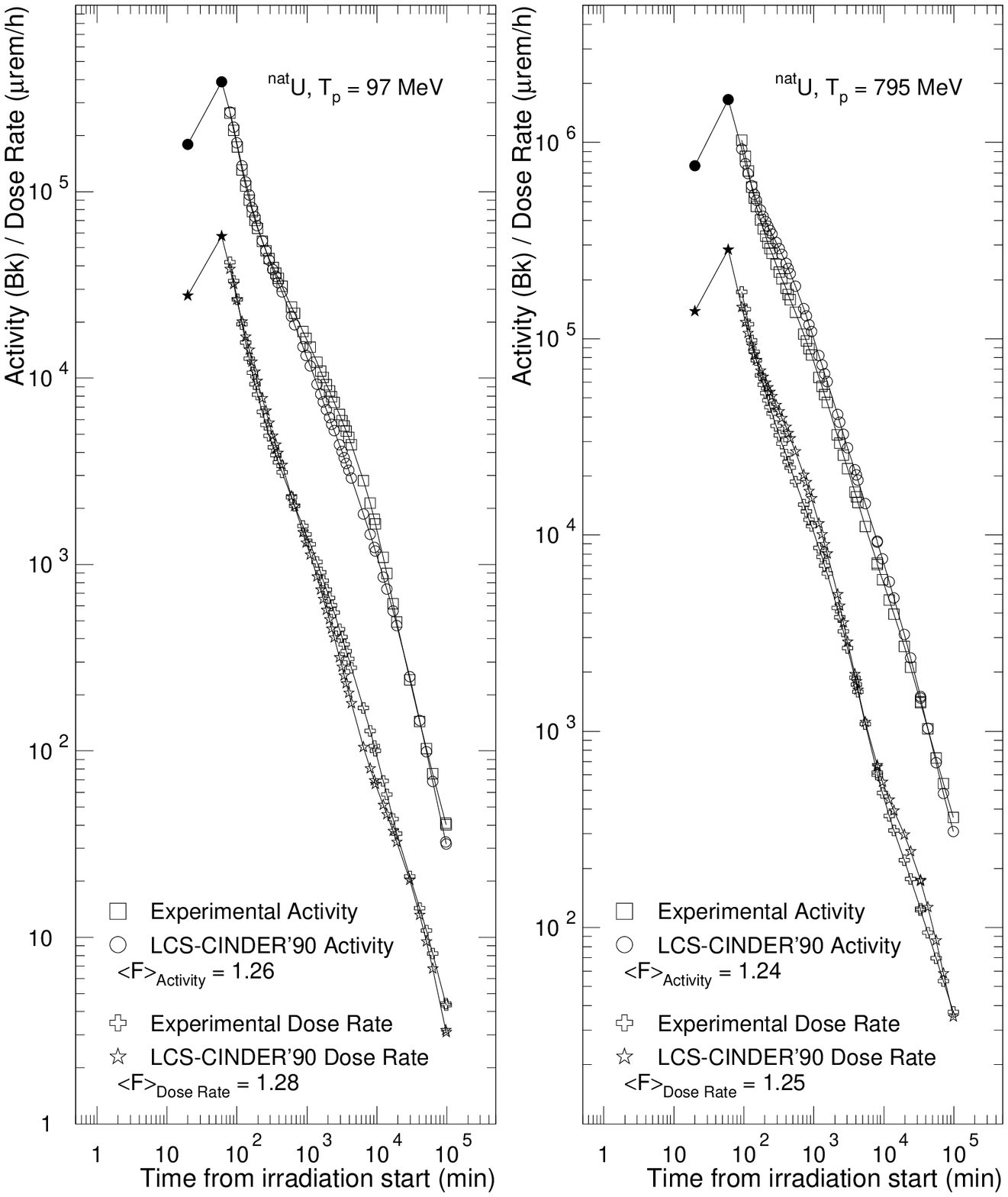}}
\caption{The same as Fig. 1, but for \nuc{nat}{U} irradiated 
with 97 MeV and 795 MeV protons.}
\end{figure*}

\end{document}